# Entropy Transfer Throughout the Structure of PDZ-2 and TIM-Barrel Proteins. A Dynamic Gaussian Network Model Study

Germán Miño-Galaz[1*], Javier Patiño Baez[2], Nicolas Miño-Berdú[3] & José González Suárez[2]

[1] Facultad de Ingenieria Ciencia y Tecnologia, Universidad Bernardo O'Higgins, Av. Viel 1497, Santiago, Región Metropolitana, Chile

[2] Departamento de Ciencias Fisicas, Facultad de Ciencias Exactas, Universidad Andres Bello, Republica 498, Santiago, Chile

[3] Departamento de Mathematica y Ciencias de la Computacion, Universidad de Santiago de Chile, Santiago, Chile

## Abstract

This research reports the entropy transfer throughout the tridimensional structure of PDZ-2 and TIM barrel structures using the dynamic Gaussian Network Model. The model predicts the allocation of the allosteric pathways of the PDZ-2. Moreover. A visualization analysis reveals that entropy and information is transported towards the effector site in PDZ-2 and near to the catalytic site of the TIM-Barrel protein. The results suggest the presence of a *functional hierarchy* that determine information and entropy flow directionality.

**Graphical Abstract**

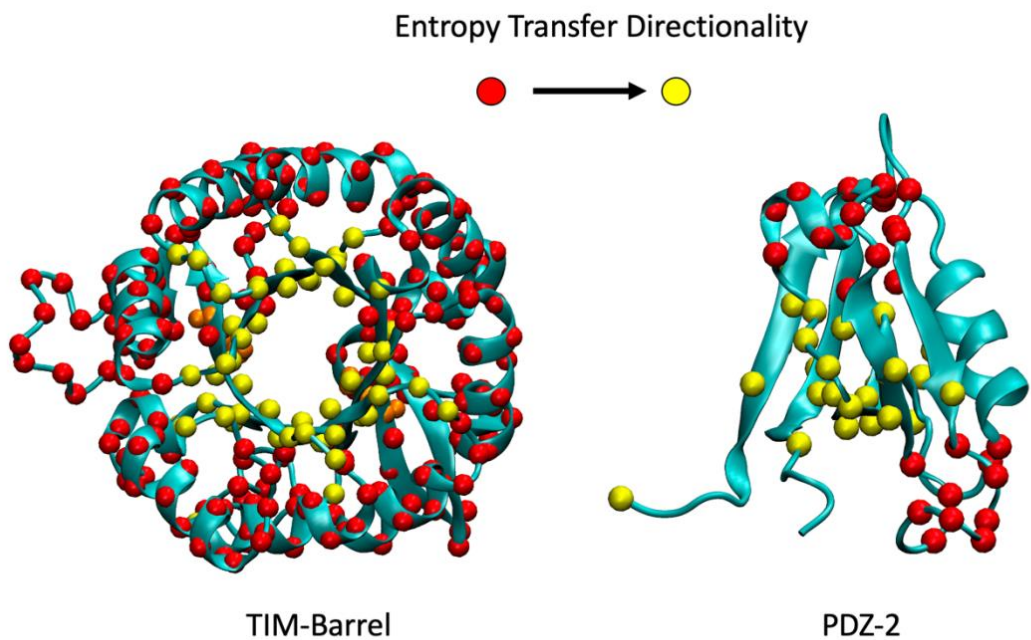

**Introduction**

Allosteric communication is a central phenomenon in molecular biochemistry. This phenomenon consists in the intramolecular information transfer throughout the three-dimensional structure of proteins. Allosteric communication requires at least two sites that interact with each other. At one hand, there is the allosteric site where an action is performed by a molecule or ligand acting to generate an input or cause; on the other hand, there is an effector site that execute a response as effect. The separation between the allosteric and the effector sites span from short to distances reaching hundreds of Å´s along the three-dimensional structure of a given protein. The modulation may consist in the increase or decrease of in the enzymatic activity or protein affinity for a given molecule or substrate at the effector site of the modulated protein. For example, negative modulation is observed when cAMP molecules bind to the catabolite activator protein (CAP). There, the binding of the initial cAMP molecule to CAP makes the binding of a second cAMP more difficult. (Popovych, 2006) Allosteric communication has three variants, one involving conformational changes that is currently named "conformational allostery", other involving almost no conformational change, named "dynamic allostery", and a third version involving a mixture of both previous forms of allostery. Conformational allosteric changes are induced by the docking of a protein and a ligand, a phenomenon firmly established by Koshland (2001). Dynamic allosteric changes have been proposed by Cooper and Dryden (1984) and is the operative mechanism of the CAM-cAMP allosteric modulation mentioned above. By their intrinsic nature, the easiest to observe is the conformational allostery, while the dynamic allostery involving a change in the vibrational properties of the system (without significant conformational change) requires a different approach. Several methods, experimental and computational, are used to study allosteric communication. (liu 2016, Chen 2022, Morales-Pastor 2022, Bernetti 2024, Nerin-Fonz 2024, Wagner 2016, Verkhivker 2020). The physical nature of the signal and the information that mediates this kind of molecular communication points to an entropy mediated phenomenon (Hacisuleyman 2017 and references wherein). This entropy transfer can be determined by several computational methods, among of them, the coarse grain dynamic Gaussian Network Model (dGNM) offers the advantage of a very low computational cost with the ability to measure relevant and precise information about intramolecular entropy transfer (Haliloglu 1997, Hacisuleyman 2017, Erkip, 2004). By measuring the spontaneous fluctuations of atoms about their average position it is possible to track entropy transfer between pair of positions in a protein system. Using the connectivity matrix between alpha carbon atoms ($C^\alpha$) of the protein, dGNM calculates vibrational normal modes and time delayed correlations to determine entropy and information transfer between pairs of $C^\alpha$ (see methods section). In this research, we study the PDZ-2 protein system and apply the dGNM and compare its prediction about allosteric pathways and entropy transfer along them, with previous allosteric characterizations of this protein (Miño-Galaz 2015, Kong 2009). This system has (at least) two well-known allosteric pathways that transfer information throughout its three-dimensional structure (depicted in Figure 1). Following Kong and Karplus (2009), two allosteric pathways that involve secondary structural elements can be distinguished. Pathway I start at strand β2 (residues 19–24) and extends along the long axis of the helix α1 (residues 44–49). Pathway II also starts at strand β2 and goes over strands β3 (residues 33–40), β4 (residues 56–61), and β6 (residues 83–90) to β1(residues 6–13). The pathways were identified by energy correlations analysis between residue pairs using classical molecular dynamics simulations (Kong 2009). So, after the ligand binds to the pocket between helix $\alpha$2 and strand β2, the signal of binding is propagated to the target regions namely, the C-terminal part on helix $\alpha$1 and the N-terminal part on strand β1 (Kong 2009). These same pathways have also been recognized as heat diffusion routes asserting an energetic role to them (Miño-Galaz, 2015). At the same time the pathways are constituted by the connectivity of the structural elements of PDZ-2 protein, so if two or more structural elements are close and connected there will be the possibility of an energy flow throughout them.

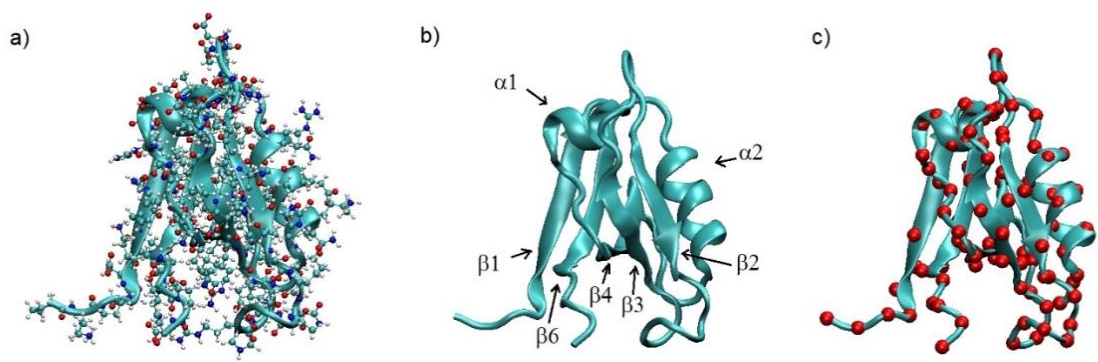

Figure 1. Model of the PDZ-2 protein (PDB ID: 3PDZ). Three representations are depicted a) all atoms superimposed with ribbon representation. b) ribbon representation depicting the labelling of the secondary structural elements, c) Ribbon representation

plus the alpha carbon atoms ($C^\alpha$) representation shown in red spheres. The lone $C^\alpha$ coordinates are used as input of the dGNM method.

PDZ-2 experiences a small conformational change in presence of its cognate ligand, thus the detection of the allosteric pathways can be done with or without its ligand. So, for simplicity, we work with the ligand free form of the protein. Thus, we report the capability of the coarse grain of dGNM to track down the pathway allocation of the PDZ-2 protein. Also, with the entropy transfer capability per $C^\alpha$ we can map upon the three-dimensional protein structure the general directionality of the signal transfer process. This mapping allows us to visualize the propagation of entropy along the protein structure. Based in this result, we propose that information transfer follows a *functional hierarchy*, in which the information is transmitted from the allosteric to the effector site. To confirm this concept, we studied the TIM-barrel scaffold. TIM-Barrel structure represents a ubiquitous scaffold that hosts at least 15 distinct enzyme families in wide range of living organisms (Wierenga 2001). The projection of entropy transfer capability of $C^\alpha\text{'s}$ in the three-dimensional structure of the TIM-Barrel shows the entropy is transferred from the peripherical hydrophilic region to central hydrophobic region of it, where the catalytic residues are located (see Figure 8). Thus, with this methodology we can map the entropy transfer towards the place where the fine tuning should be exerted, namely, the active or catalytic site of a system.

**Theoretical methods and computational details.**

The dGNM model for a folded protein assigns interaction of close residues by linear springs, thus the protein is analyzed in the context of a three-dimensional elastic network. The junctions or nodes are the alpha carbon atoms ($C^\alpha$) of the protein. The position of the i-th $C^\alpha$ is denoted by $R_i$ and its gaussian fluctuation around the equilibrium position is expressed as $\Delta R_i(t) = R_i - \overline{R}_i$ where $\overline{R}_i$ is the mean position. The dGNM method uses a connectivity matrix $\Gamma$ that is defined as:

$$\Gamma_{ij} = \begin{cases} -1 & \text{if } i \neq j \text{ and } R_{ij} \leq r_c, \\ 0 & \text{if } i \neq j \text{ and } R_{ij} > r_c, \\ -\sum_{i, i \neq j} \Gamma_{ij} & \text{if } i = j. \end{cases}$$

Equation 1.

Here $R_{ij}$ is the distance between the $i^{th}$ and $j^{th}$ atoms, and $r_c$ is a cut-off distance of 7.0 A (in our case). The diagonal elements are the negative sum of the off-diagonal elements of the $i^{th}$ column. So, $\Gamma$ is an n x n symmetric matrix, where n is the number of residues. $\Gamma$ is also known as local packing or coordination density matrix. The diagonalization of $\Gamma$ is expressed as $\Gamma^{-1} = U(\Lambda)^{-1}U^T$, where U is an orthogonal matrix with columns of the eigenvectors $u_i$ of the matrix $\Gamma$, and $\Lambda$ is a diagonal matrix composed by the eigenvalues $\lambda_i$ of $\Gamma$. It is possible to decompose $\Gamma^{-1}$ as the sum of contributions from individual modes as

$$\Gamma^{-1} = \sum_{k=2}^{n} \lambda_k^{-1} u_k u_k^T = \sum_{k=2}^{n} A^{(k)}$$

Equation 2.

Where $A_{ij}^{(k)}$ is an n x n matrix that describes the contribution of the k-th vibrational mode to the atomic fluctuations, thus $k=2$. The first eigenvalue of $\Gamma$ is cero is not included in the summation. The time delayed correlation between the $i^{th}$ and $j^{th}$ $C^\alpha$ atoms can be expressed as the sum of the contributions of individual modes as

$$\langle \Delta \mathbf{R_i}(0) \Delta \mathbf{R_j}(\tau) \rangle = \sum_k \mathbf{A_{ij}}^{(k)} exp\left(\frac{-\lambda_k \tau}{\tau_0}\right)$$

Equation 3.

Here $\tau$ is the time delay, and $\tau_0$ is a characteristic time usually of 6 ps for proteins (full details of the solution leading to equations 1, 2 and 3 are shown in Erkip, 2004). Applying Schreiber's work in information transfer (Schreiber 2000), Erman describes a way to quantify the time delayed information transfer between pairs of residues in a protein (Hacisuleyman, 2017). So, information transfer depends on conditional probabilities evaluated for the fluctuations of residues i and j. The fluctuation of the two-vector position $\Delta R_i(t)$ and $\Delta R_j(t)$ change rapidly with time. If the fluctuation of the two nodes is correlated, the knowledge of the fluctuation of the $i^{th}$ node (at t = 0) will decrease the uncertainty of the knowledge (at t = $\tau$) of the fluctuations of the $j^{th}$. The time dependence of two correlations is given by $\langle \Delta \mathbf{R_i}(0) \Delta \mathbf{R_j}(\tau) \rangle$, given by equation 3. If the asymmetric relation $\langle \Delta \mathbf{R_i}(0) \Delta \mathbf{R_j}(\tau) \rangle \neq \langle \Delta \mathbf{R_j}(0) \Delta \mathbf{R_i}(\tau) \rangle$ holds, a net information transfer from one residue to another is occurring. Thus, the entropy transfer $T_{i \rightarrow j}(\tau)$ from the trajectory $\Delta R_i(t)$ of residue i to the trajectory $\Delta R_j(t)$ of residue J is the amount of uncertainty loss in future values of $\Delta R_j(t + \tau)$ by knowing the past values of $\Delta R_j(t)$.

The working equation, obtained by Erman, to determine the entropy transfer between two protein nodes is:

$$T_{i \to j}(\tau) = \frac{1}{2} \ln \left( \left( \sum_k A_{jj}^{(k)} \right)^2 - \left( \sum_k A_{jj}^{(k)} \exp\{-\lambda_k \tau/\tau_0\} \right)^2 \right)$$

$$- \frac{1}{2} \ln \left[ \left( \sum_k A_{ii}^{(k)} \right) \left( \sum_k A_{jj}^{(k)} \right)^2 \right.$$

$$+ 2 \left( \sum_k A_{ij}^{(k)} \right) \sum_k A_{jj}^{(k)} \exp\{-\lambda_k \tau/\tau_0\} \sum_k A_{ij}^{(k)} \exp\{-\lambda_k \tau/\tau_0\}$$

$$- \left\{ \left( \sum_k A_{ij}^{(k)} \exp\{-\lambda_k \tau/\tau_0\} \right)^2 + \left( \sum_k A_{ij}^{(k)} \right)^2 \right\} \left( \sum_k A_{jj}^{(k)} \right)$$

$$\left. - \left( \sum_k A_{jj}^{(k)} \exp\{-\lambda_k \tau/\tau_0\} \right)^2 \left( \sum_k A_{ii}^{(k)} \right) \right] - \frac{1}{2} \ln \left[ \left( \sum_k A_{jj}^{(k)} \right) \right]$$

$$+ \frac{1}{2} \ln \left( \left( \sum_k A_{ii}^{(k)} \right) \left( \sum_k A_{jj}^{(k)} \right) - \left( \sum_k A_{ij}^{(k)} \right)^2 \right)$$

Equation 4.

With the elements of matrix A playing a central role in it. Again $\tau$ is the time delay, and $\tau_0$ is a characteristic time as is shown for Equation 3 are defined above. The solution leading to Equation 4 it fully shown in Hacisuleyman 2017 and in its supplementary material. We used equation 4 to analyze of the allosteric pathways of the PDZ-2 system and to compare its results with previous ones (Miño-Galaz 2015). So, we calculated the information transfer based on Equation 4 using a python-based algorithm. The calculation is done using on the first NMR data of the protein data bank (PDB) structure 3PDZ. The algorithm extracts the coordinates of the $C^\alpha$ form the PDB file, constructs the $\boldsymbol{\Gamma}$ matrix, determines the normal mode vector and eigenvalues to calculate the inverse $\boldsymbol{\Gamma}^{-1}$. As $\boldsymbol{\Gamma}^{-1} = \sum_{k=2}^{n} A^{(k)}$, its elements are used to calculate the residue-to-residue entropy transfer. The output for each protein residue is used to build a level curve for entropy transfer as is shown in Figure 2a. A secondary characterization is based on the information transfer for one node to the rest of the protein using the summation

$$T_{i\to\odot}(\tau) = \sum_j T_{i\to j}(\tau)$$

Equation 5.

Where $T_{i\to\odot}(\tau)$ represents the total information that the i[th] element transfers to the rest of the nodes ($C^\alpha$) of the protein at the time delay $\tau$.

**Results.**

Figure 2a shows the results of dGNM entropy transfer simulation for $\tau$ value of 0.2. Strong similarity between the dGNM transfer map with its respective connectivity map is observed (Figure 2b). This fact reveals that the transfer is mediated by the proximity of the structural components of the network that compose the allosteric pathways. From the transfer map, the cognate allosteric pathways I and II of the PDZ-2 system can be readily observed. For Pathway I, which involves the allosteric information transfer between the secondary structure elements β2 and α1 an entropy transfer is directly observable (marked with an ellipse). For Pathway II the coupling between elements that gives rise to it can be observed by the sequential organization of the transfer map, namely, β2 → β3 → β4 → β6 → β1. The sequential path is denoted with squares figures numbered from 1 to 4. The square 1 denotes the interaction between β2 → β3; square 2 the transference between β3 → β4; square 3 denotes the coupling β4 → β6; and square 4 denotes the β6 → β1 interaction. The coupling between all the elements of this pathway can be also observed by information contained in form each of the emitting residues that compose the entropy transfer map in Figure 2a. In line with this, Figure 3a shows the entropy transmission of emitting position 19 to the rest of the structure. This position is located in the β2 (positions 19 to 24) structural element and transmits a strong signal to positions 40 to 55, that contain the α1 structural element, which conforms to Pathway I (peaking around position 44). This signal of nearly perfectly mirrored by emitting position 47 (peaking position 19) as is shown in figure 3b. This pair of signals show an asymmetrical feature. This means that a given position may not receive all the entropy that transmits to another position, which is an expected feature of the directionality of the allosteric phenomena. In this case, position 19 transmits 0.0348 entropy units to position 47, whereas position 47 transmits 0.0427 units to back position 19. Similar observations can be made for Pathway II (Figure 4). Observing emitting position 23 the direct propagation of the entropic signal can be traced out to signals in the β3, β4, β6 & β1 secondary structure elements which are part of this pathway. These signals peak at positions 36, 55, 86 & 11 respectively (Figure 4a). The entropy transfer along this pathway can be traced out from element 36 to the other components of it. A similar transmitting capacity is observed for other emitting elements that belong to this same pathway (data not shown). So, the entropic signals tend to be transmitted to the elements of the pathway to which the emitter belongs. For this same Pathway II, a nonreciprocal relationship can be also observed. In this case, position 23 transmits 0.0115 entropy units to position 36 (Figure 4a), whereas position 36 transmits only 0.0086 units to position 23 (Figure 4b). The observed asymmetric directionality in signal propagation has been observed in other reports (Miño-Galaz, 2015) and is representative, to our understanding, of the directional nature of allosteric communication.

It is also worth mentioning the distance that an entropy signal can travel, spanning a wide range of distances. For instance, in one case the position 19 (Pathway I) transmits entropy to position 47 which involves a rough distance of 6 Å. In the case of Pathway II, the transmitting position 23 propagates a signal that involves nearly 30 Å to position 86 along the contour of this pathway. Its seem that $\beta 2$ sheet (residues 19 to 23) is a hub for information transfer in the protein. Finally, In the case Pathway II new signal is also observed coming from the loop connecting $\beta 2$ and $\beta 3$, peaking at positions 71 and 70 in Figure 4a and 4b, respectively. The detection of this new spot is consistent with previous report about the allocation of allosteric pathways in the PDZ-2 system using Nuclear Magnetic Resonance (NMR) methods. The usage nuclear Overhauser effect (Ashkinadze 2022) and methyl relaxation data (Fuentes 2004) demonstrate that this loop is part of allosteric network of the PDZ-2 protein.

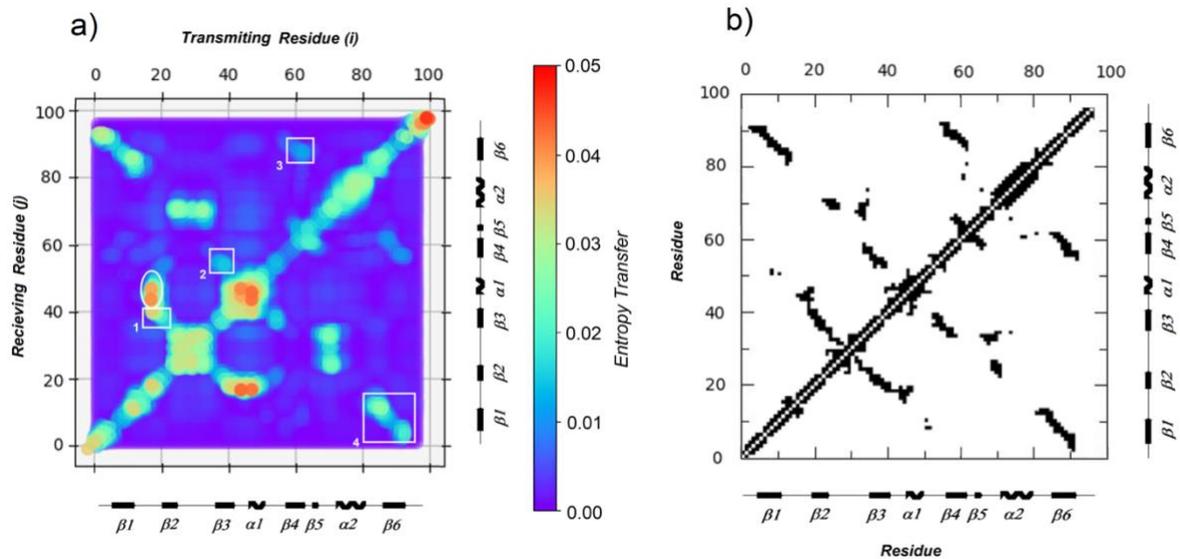

Figure 2. a) Entropy transfer map for PDZ-2 using dGNM. Elliptic figure highlight Pathway I and the square boxes denoted Pathway II (see text for further details) b) Contact map of PDZ-2 calculated using Prody-GNM algorithm. Other allosteric couplings can be observed in the entropy transfer map apart from Pathways I & II. The contact map is a binary one, meaning that contacts below the cutoff of 7.0 A are in black and contacts above are in white.

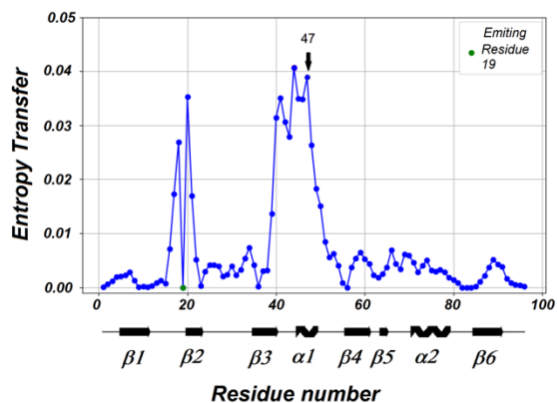
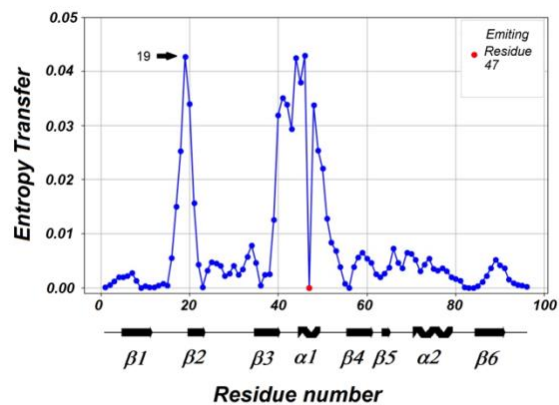

Figure 3. a) Entropy transfer for emitting residue 19 to re rest of the protein (Left).

b) Entropy transfer for emitting residue 47 to re rest of the protein (Right)

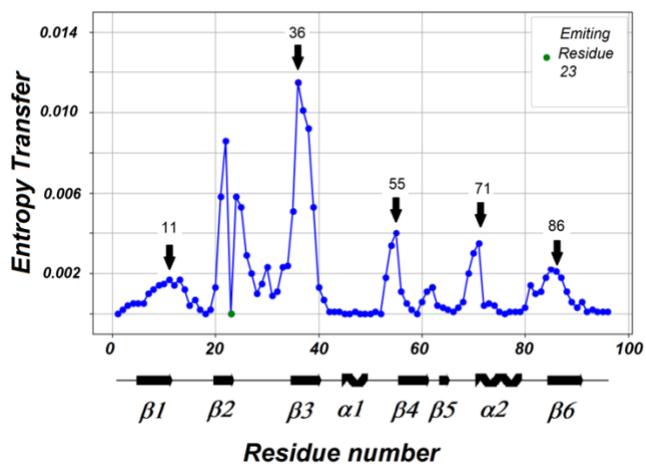
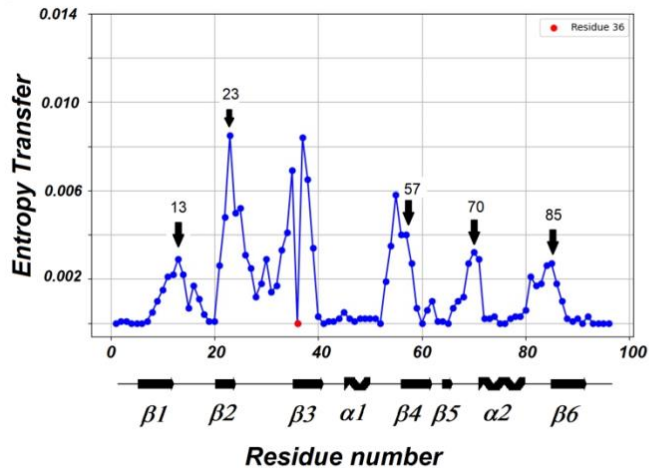

Figure 4. a) Entropy transfer for emitting residue 23 to re rest of the protein (Left).

b) Entropy transfer for emitting residue 36 to re rest of the protein (Right)

These results are consistent with previous analysis of the allosteric pathways of the PDZ-2 system. Using an all-atom non-equilibrium method called anisotropic thermal diffusion (ATD) it was previously shown that the PDZ-2 allosteric pathways are linked to the heat diffusion pathways, and that heat diffusion pathways are determined by the connectivity of structural elements of the protein (Miño-Galaz 2015). The result of that previous study is depicted in Figure 5. In that study the protein structure without solvent is cooled to 10 K then each amino acid was heated by separated to 300 K. This procedure revealed that the heat flows throughout the protein contact network which at the same time are the cognate allosteric pathways of the PDZ-2 system. That report was also consistent with the correlated movements analysis prior done by Kong (2007), which is the initial reference to define allosteric Pathways I and II. Now in this actual dGNM report we found an entropy transfer runs through the contact network the protein and defines the cognate allosteric pathways of this system, which involves the structural elements $\beta 2 \rightarrow \alpha 1$, for Pathway I and $\beta 2 \rightarrow \beta 3 \rightarrow \beta 4 \rightarrow \beta 6 \rightarrow \beta 1$ for Pathway II. All these observations confirms that the dGNM method, despite being a coarse grain model, is a sensitive and precise method to characterize the allocation of the allosteric communication pathways in protein systems.

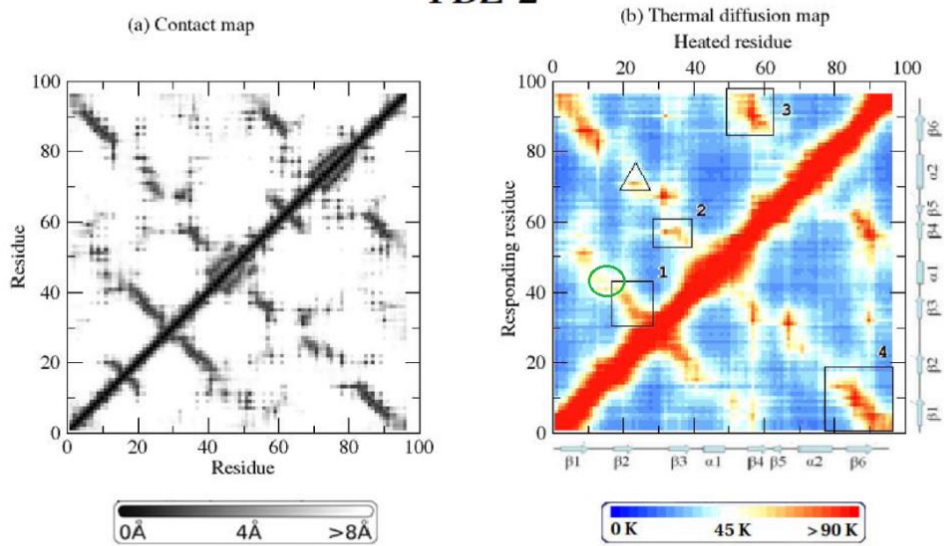

Figure 5.- Previous ATD study for the PDZ-2 system. Elliptic figure highlight Pathway I (β2 → α1) and the square boxes 1 to 4 denoted pathway II (β2 → β3 → β4 → β6 → β1). Other allosteric couplings can be observed in the thermal diffusion map apart that may not belong to Pathway I or II. (Reprinted and adapted with permission from *The Journal of Physical Chemistry B*, *119*(20), 6179-6189. Copyright 2015 American Chemical Society)

Next, we decided to explore the distribution of the capability of a given $C^\alpha$ to propagate allosteric information. For this we use equation 5, which is the simple summation of the entropy transfer from one node to the rest of the molecule. So, we obtain the transfer capability of each node. Then, we ranked it the $C^\alpha$ according to this transfer capability and inspect its allocation upon the three-dimensional structure of the PDZ-2 protein. From this ranking, we took the 27 best donors and the 27 worst donors, to highlight its allocation upon the PDZ-2 structure (Figure 7). The $C^\alpha$ with intermediary transfer capabilities are not shown for simplicity, but their allocation can be easily deduced from the inspection of Figure 7. According to this analysis the best information receiving site is the structural beta sheet that contains β3, β4, β6 and β1, while the best information emitting sites are α1 helix and the loop connecting the β2 and β3 (loop β2 - β3 in figure 7a) which is close to the ligand binding site at the pocket between helix α2 and strand β2. Despite to being gross best/worst information transfer visualization this organization of is consistent with the overall directionality of pathways I and II, with β1 as an ending point. This simple evaluation made us think about the presence of *functional hierarchy* that determines the allosteric information transfer from a site to another site of the protein. This *functional hierarchy* concept is nothing else than the general paradigm of allosteric communication, that asserts that information is transferred from effector site to the allosteric site. Here, this hierarchy is three-dimensionally mapped out from the entropy transfer capability of each $C^\alpha$. According to our understanding this functional hierarchy is inscribed in the three-dimensional a structure of the analyzed systems.

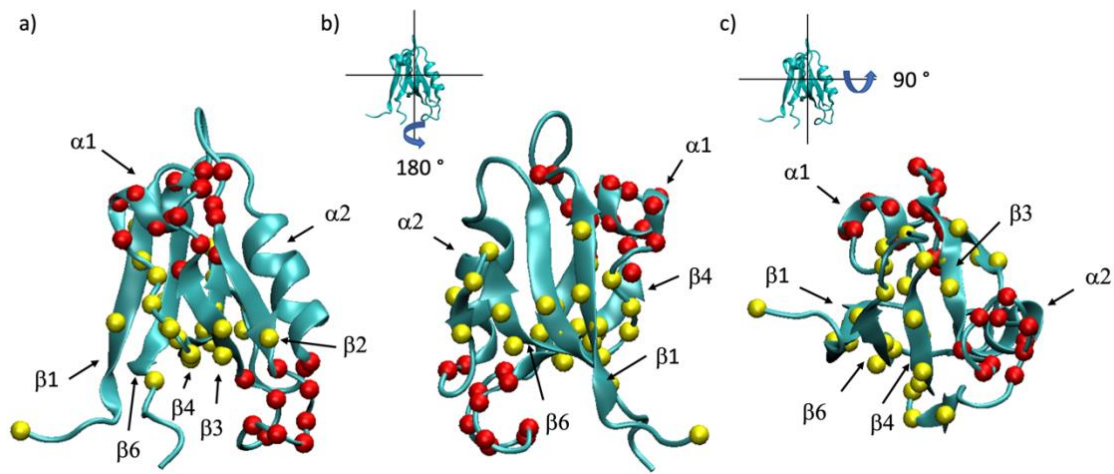

Figure 7. The 27 best entropy donors (red spheres) and the 27 worst entropy donors (yellow spheres) highlighted upon the PDZ-2 structure. a) Front reference view, b) Back side view : 180° rotation with respect to the vertical axis, c) Bottom view : 90° rotation with respect to the horizontal axis.

Then, we decided to test this concept on a higher scale by inspecting the alpha/beta barrel protein structure, also generically known as TIM-Barrel. For this case we analyzed the TIM-Barrel protein HisF-C9S (PDBid 5TQL) composed 252 amino-acids residues. Thus, we pass its structure through the dGNM algorithm (using Equation 4), and generate the list of best/worst information donors (using Equation 5) and allocate them upon the three-dimensional structure of the barrel. Figure 8a shows the new cartoon backbone structure of the TIM-Barrel depicting in orange (as reference) the residues Asp130 and Asp11 which are part of the catalytic residues of this system (Beismann-Driemeyer 2001) and the 196 best (red spheres) and the 56 worst (yellow spheres) donors allocated upon the TIM-Barrel Structure. Figure 8c shows the backbone structure of the protein with the residues Asp130 and Asp11 for best visualization of them. Figure 8d shows sagittal view with the allocation of the worst 56 information donors with Asp130 and Asp11 residues. Here is observed the existence of this *functional hierarchy* of the best/worst information donors, with the central region harboring the worst donors, while the peripherical region containing the best information donors. It's clear from the visualization that the information goes from the outer regions of the protein to the inner regions of it, especially to the central region that contains the catalytic residues. Also, the three-dimensional projection suggests that the information is transferred mainly to the hydrophobic regions of the protein. It is known that this TIM-Barrel structure has a lower N-terminal region of the $\beta$-sheet ($\alpha\beta$ loop region) that is dedicated to give stability to the protein, and a C-terminal region of the $\beta$-sheet ($\beta\alpha$ loop region) that is dedicated to hold the catalytic activity (Wierenga 2001). Our analysis of best/worst donor clearly shows that the information is transported where it is most needed, that it is very close to the region containing the catalytic residues. It is important to mention that the TIM-Barrel structure represents a ubiquitous scaffold that hosts at least 15 distinct enzyme families in wide range of living organisms (Wierenga 2001). Also, the TIM-Barrel general organization is actively used for de novo enzymatic design and engineering (Romero-Romero 2021). Finally, allosteric modulation has been reported for this kind of biological macrostructure

(Gamiz-Arco 2021, Chan 2020). This result support the existence of functional *hierarchy* that dominates the allosteric information transfer directionality. Further research is needed to unveil the topological and dynamical features that underpin this hierarchy and the entropy transfer capability of different protein regions.

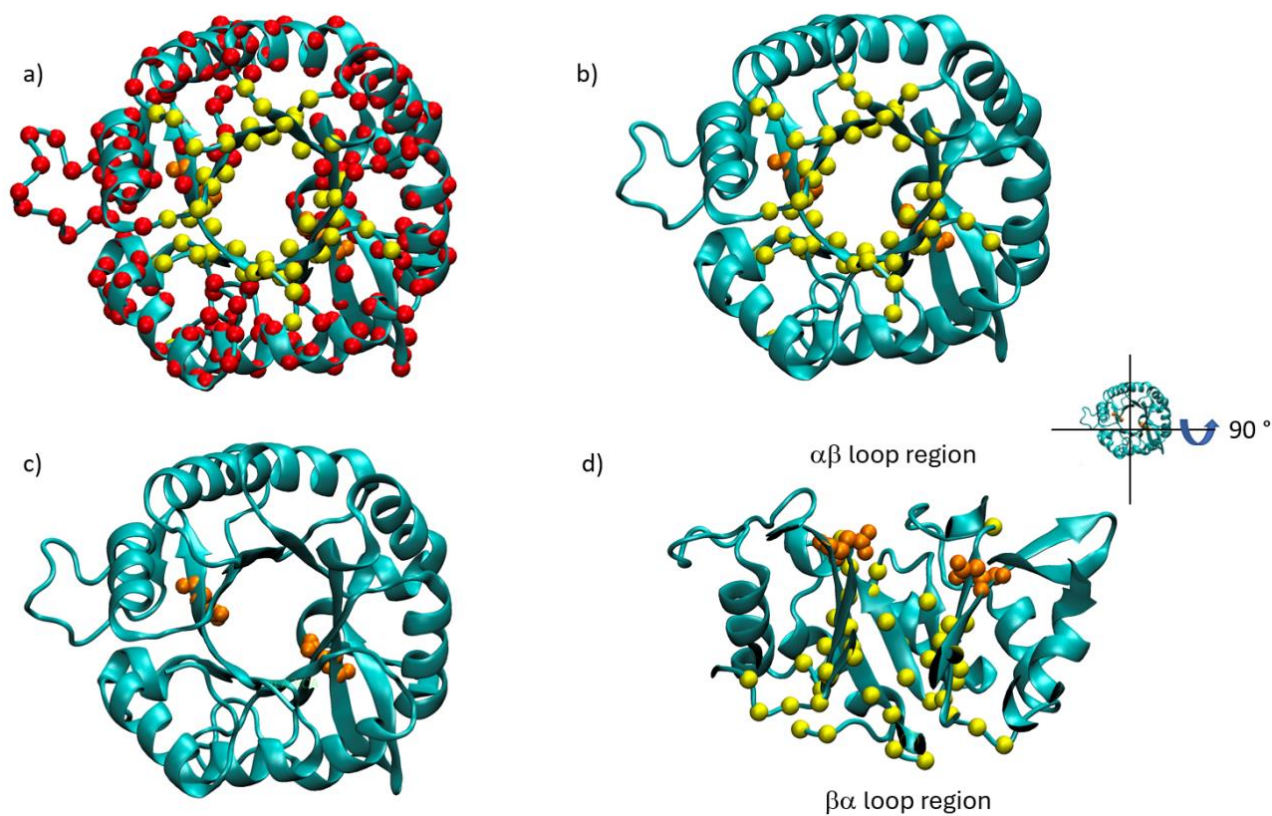

Figure 8. Allocation of the best/worst donors upon the 3-dimensional structure of the TIM-Barrel protein HisF-C9S (PDB id 5TQL). a) Cartoon diagram of the protein highlighting the catalytic residues Asp130 and Asp11 in orange (as reference) with 194 best (red spheres) and 56 worst (yellow spheres) entropy donors allocated upon the structure b) view of the catalytic residues Asp130 an Asp11 and the 56 worst (yellow

spheres) allocated upon the structure c) view of the catalytic residues Asp130 an Asp11 upon the TIM-Barrel. d) Sagittal view of the TIM-barrel showing that allocation of the worst donors with respect to the catalytic residues. The worst donors are allocated in stabilization region ($\beta\alpha$ loop) next to the catalytic region ($\alpha\beta$ loop).

**Discussion and Conclusion.**

In this article we explored the capabilities of the dGNM to predict the allocation of allosteric pathways in a well-known system, namely, PDZ-2, by measuring the entropy transfer along the three-dimensional structure of the protein. Despite being a coarse-grained model, dGNM it accurately revealed the presence of the known allosteric pathways of the PDZ-2 system and its details. This conclusion is supported by direct comparison with previous heat transfer studies. Also, a particular usage of the method did lead to hypothesizing the presence of a *functional hierarchy* for the information transfer capability. This effect is clearly mapped by plotting the best and worst information donors upon the three-dimensional structure PDZ-2 system. This functional hierarchy just implies that information is transferred from recruiting or allosteric sites towards an effector site of the proteins. This hierarchy was measured in much more complex macromolecular protein system, namely, the TIM-Barrel scaffold. The result showed that the allocation of the best and worst information donors responds to the correct functionality of the scaffold, meaning the the information is transferred to the regions that contain the catalytic residues. This protein scaffold is a common motive that accommodates several and different enzyme activities along nature. Both analyzed systems strongly suggest the presence of this *functional hierarchy* in which the information is transferred to the site of the molecules where delicate modulation it is mostly needed. The regulation of the catalytic site requires information, and this information is provided by a delicately established allosteric network. Although we are working with a coarse grain model, this description shows to be completely consistent with the known data for the allosteric pathways of the PDZ-2 system and with the biochemical nature of the alpha/beta barrel (or TIM barrel) in which the catalytic activity is allocated in the core of this well know enzymatic system. In general we see dGNM as a first approach method to characterize allosterism in proteins in consistency with previous literature. In particular, the allocation of entropy tranfer capability present us an oportunity to get a best understanding of the overall allosteric process.

**Statement:** During the preparation of this work no generative AI was used.

**Acknowledgements**: J.M. Gonzalez and G. Miño-Galaz are thankful UNAB for a grant DI-17-20/REG-VRID. Powered@NLHPC: This research/thesis was partially supported by the supercomputing infrastructure of the NLHPC (CCSS210001)